\date{}
\documentclass[showpacs]{revtex4}
\usepackage{graphicx,psfrag,amsmath,amssymb,amsfonts,bbm,latexsym,color,dcolumn}
\begin{document}
\title{Casimir effect with dynamical matter on thin mirrors}
\author{C. D. Fosco$^a$ }
\author{F. C. Lombardo$^b$ }
\author{F. D. Mazzitelli$^b$ }
\affiliation{$^a$Centro At\'omico Bariloche and Instituto Balseiro,
 Comisi\'on Nacional de Energ\'\i a At\'omica, \\
R8402AGP Bariloche, Argentina}
\affiliation{$^b$Departamento de
F\'\i sica {\it Juan Jos\'e Giambiagi}, FCEyN UBA, Facultad de
Ciencias Exactas y Naturales, Ciudad Universitaria, Pabell\' on I,
1428 Buenos Aires, Argentina.}
\date{today}

%====================================================================

\begin{abstract}
\noindent We calculate the Casimir energy for scalar and gauge fields in
interaction with zero-width mirrors, including quantum effects due to the
matter fields inside the mirrors.  We consider models where those fields
are either scalar or fermionic, obtaining general expressions for the
energy as a function of the vacuum field 1PI function.  We also study,
within the frame of a concrete model, the role of the dissipation induced
by those degrees of freedom, showing that, after integration of the matter
fields, the effective theory for the electromagnetic field contains modes
with complex energies. As for the case of Lifshitz formula, we show that
the formal result obtained by neglecting dissipation coincides with the
correct result that comes from the quantum fluctuations of both  bulk and
matter fields.
\end{abstract}
%\pacs{}
\maketitle
%%%%%%%%%%%%%%%%%%%%%%%%%%%%%%%%%%%%%%%%%%%%%%%%%%%%%%%%%%%%%%%%%%%%%
%%%%%%%%%%%%%%%%%%%%%%%%%%%%%%%%%%%%%%%%%%%%%%%%%%%%%%%%%%%%%%%%%%%%%
%%%%%%%%%%%%%%%%%%%%%%%%%%%%%%%%%%%%%%%%%%%%%%%%%%%%%%%%%%%%%%%%%%%%%
%%%%%%%%%%%%%%%%%%%%%%%% Introduction %%%%%%%%%%%%%%%%%%%%%%%%%%%%%%%
%%%%%%%%%%%%%%%%%%%%%%%%%%%%%%%%%%%%%%%%%%%%%%%%%%%%%%%%%%%%%%%%%%%%%
%%%%%%%%%%%%%%%%%%%%%%%%%%%%%%%%%%%%%%%%%%%%%%%%%%%%%%%%%%%%%%%%%%%%%
%%%%%%%%%%%%%%%%%%%%%%%%%%%%%%%%%%%%%%%%%%%%%%%%%%%%%%%%%%%%%%%%%%%%%
It has become  increasingly clear that the use of sensible models for the
description of the media that constitute the mirrors is an unavoidable step
for the refinement of Casimir energy calculations~\cite{rev}.  In
particular,  the theory of quantum open systems is the natural approach to
clarify  the role of dissipation in Casimir physics.  In this framework,
dissipation and noise appear in the effective theory of the relevant
degrees of freedom (the electromagnetic field) after integration of the
matter degrees of freedom.

In this letter we consider mirrors described by thin films, equipped with
matter fields confined to them; these localized fields shall induce an
effective action for the coupling of the vacuum fields to the mirrors.  It
will turn out that the resulting effective action falls into a class of
model that we had studied in a previous work~\cite{flm08}.  We then apply
those results here in a quite straightforward way, obtaining the Casimir
energy for different cases: regarding the vacuum fields, we shall consider
the cases of a massless (real) scalar field and of an Abelian gauge field.
As we shall see, under our general assumptions, the Casimir energy of the
latter may be obtained as a derived result of the former, after performing
some identifications.

The Casimir effect produced by thin plasma
sheets has been previously considered by other authors (see~\cite{bordag06,bordag07}
and references therein)  imposing appropriate
boundary conditions on the position of the sheets~\cite{bordag06},
considering a fluid of non relativistic electrons confined to the
surface~\cite{bordag07}, or describing the matter inside thin films using
the particle in a box model \cite{benassi}.   We will extend these results to the case of
relativistic bosonic or fermionic degrees of freedom confined to the
mirrors. Besides, for a particular fermionic model, we will
show that, in the effective theory for the electromagnetic field, the
modes have complex energies, due to the possibility of transferring  energy
from the bulk to the thin surfaces. The Casimir energy is of course
always real, and given by the sum of eigenfrequencies of the full system
(bulk and matter fields).

%%%%%%%%%%%%%%%%%%%%%%%%%%%%%%%%%%%%%%%%%%%%%%%%%%%%%%%%%%%%%%%%%%%%%
%%%%%%%%%%%%%%%%%%%%%%%%%%%%%%%%%%%%%%%%%%%%%%%%%%%%%%%%%%%%%%%%%%%%%
%%%%%%%%%%%%%%%%%%%%%%%%%%%%%%%%%%%%%%%%%%%%%%%%%%%%%%%%%%%%%%%%%%%%%
%%%%%%%%%%%%%%%%%%%%%%%% Effective action %%%%%%%%%%%%%%%%%%%%%%%%%%%
%%%%%%%%%%%%%%%%%%%%%%%%%%%%%%%%%%%%%%%%%%%%%%%%%%%%%%%%%%%%%%%%%%%%%
%%%%%%%%%%%%%%%%%%%%%%%%%%%%%%%%%%%%%%%%%%%%%%%%%%%%%%%%%%%%%%%%%%%%%
%%%%%%%%%%%%%%%%%%%%%%%%%%%%%%%%%%%%%%%%%%%%%%%%%%%%%%%%%%%%%%%%%%%%%

Let us first consider the effective action, $S_{eff}^{(I)}$, for a real
scalar field $\varphi$, which results from its interaction with matter
confined to two identical, flat, parallel and zero-width mirrors in $d$
spatial dimensions, defined by the equations $x_d = 0$ and $x_d=a$.  The
total Euclidean action $S$ for $\varphi$ will then be of the form:
\begin{equation}\label{eq:defs}
S(\varphi) \;=\; S_0(\varphi) \,+\, S_{eff}^{(I)}(\varphi) \;,
\end{equation}
where $S_0 \equiv \frac{1}{2}\int d^{d+1}x \, \partial_\mu \varphi
\partial_\mu \varphi$,  defines the free theory. $S_{eff}^{(I)}$ do,
of course, depends on the nature of
the fields confined to the mirrors and on the structure of their
interaction with $\varphi$. The functional form of $S_{eff}^{(I)}$ will be
assumed to be quadratic in $\varphi$. This  approximation
is justified by the following reason: since the media are
assumed to impose (approximate) Dirichlet boundary conditions, the relevant
configurations correspond to {\em small\/} values for the fields on the
mirrors (otherwise there would not be approximate boundary conditions
there).  It is then legitimate to assume that higher-order terms in those
fields shall be strongly suppressed in comparison with the lowest, non
trivial one (quadratic), since they will have a smaller weight in the
functional integral.

\noindent With this remark in mind, taking into account the fact that the
matter fields are confined to the mirrors, and recalling that the
interaction is assumed to be local, we have:
\begin{equation}
S_{eff}^{(I)}(\varphi) \,=\, S_{eff}^{(1)}(\varphi)
\,+\,S_{eff}^{(2)}(\varphi)
 \end{equation}
where:
$$
S_{eff}^{(\alpha)}(\varphi)\,=\,
\frac{1}{2}\int dx_0 \int dx'_0 \int d^{d - 1}x_\parallel
\int d^{d-1}x'_\parallel \int dx_d \int dx'_d
$$
\begin{equation}\label{eq:defsia}
\times \varphi(x_0,{\mathbf x}_\parallel, x_d) \,
\lambda(x_0-x_0'; {\mathbf x}_\parallel - {\mathbf
x}_\parallel') \delta(x_d - a_\alpha)\delta(x'_d - a_\alpha) \,
\varphi(x'_0,{\mathbf x}_\parallel ', x_d)\;,
\end{equation}
with $a_1\equiv 0$ and $a_2\equiv a$. ${\mathbf x}_\parallel$ denotes the
$d-1$ coordinates parallel to the mirror: $x_1,x_2,\ldots, x_{d-1}$.

The interaction term will be obtained as an effective action coming from
the integration of the degrees of freedom confined to the walls that
interact {\em locally\/} with the scalar field $\varphi(x)$ at $x_d=0$ and
$x_d=a$.  To simplify the notation in the following calculations, we shall
denote by $y_a$ the spacetime coordinates for the
matter fields, with $y_a \equiv x_a$, for
$a=0,1,\ldots,d-1$.

Using a Fourier transformation with respect to the $y_a$ coordinates,
\begin{equation}\label{eq:siaf}
S_{eff}^{(\alpha)}(\varphi)\,=\,
\frac{1}{2}\int \frac{d^dk}{(2\pi)^d}\,
{\tilde\varphi}^*(k,a_\alpha) \,{\tilde\lambda}(k)
\, {\tilde\varphi}(k,a_\alpha)
\end{equation}
where $k \equiv (k_a)_{a=0}^{d-1}$, are the Fourier space variables
associated to the mirror's spacetime coordinates.  Thus, the effect of the
matter modes can, for this case, be encoded in a single function
${\tilde\lambda}(k)$.  Moreover, depending on whether the dynamics of the
matter fields is assumed to be relativistic or non relativistic, it will
depend on $\kappa \equiv \sqrt{k_a k_a}$, or on $k_0$ and $|{\mathbf
k}_\parallel|$, respectively.

We see that the term (\ref{eq:siaf}) is the $\epsilon \to 0$ version of the
models we have introduced in~\cite{flm08}. Thus, the result of the derivation
presented there for the Casimir energy has direct applicability here. That is
true, regardless of the microscopic model assumed for the matter field;
however, before writing the expression for the energy, let us derive an
expression for $\tilde{\lambda}$ from a concrete model. Note that, since the
structure of $S_{eff}^{(\alpha)}$ depends trivially on the value of $a_\alpha$,
we just need to consider the $\alpha=1$ case, shifting the field afterwards.

A first model emerges from the assumption that $S_{eff}^{(1)} \equiv S_{eff}$ is due to
the linear coupling of $\varphi$ to a microscopic real scalar field $\xi(y)$, in $d$ spacetime
dimensions, equipped with an action $S_m(\xi)$:
\begin{equation}\label{eq:deseff}
e^{-S_{eff}(\varphi)} \;=\; \frac{\int {\mathcal D}\xi \, e^{-S_m(\xi)
\,+\, i g \int d^dy \, \xi(y) \, \varphi(y,0)}}{\int {\mathcal D}\xi \,
e^{-S_m(\xi)}} \;,
\end{equation}
where $g$ is a coupling constant. The matter field $\xi$ may have a
self-interaction, controlled by an independent coupling constant, implicit
in $S_m$.

To proceed, we denote by $W(J)$ the generating functional of connected correlation
functions of $\xi$ (in a $d$-dimensional spacetime), related to ${\mathcal
Z}(J)$, the one for the full correlation functions:
\begin{equation}
{\mathcal Z}(J)\,=\, \int {\mathcal D}\xi \, e^{-S_m(\xi) \,+\,\int d^dy
J(y) \xi(y)} \;,
\end{equation}
by $W = \ln {\mathcal Z}$. We then have that $S_{eff}(\varphi) = -
W[ i \,g\, \varphi(y,0)]$.
On the other hand, since only the quadratic part will be retained,
\begin{eqnarray}
S_{eff}(\varphi) &=& -  W\big[i \,g\, \varphi(y,0)\big] \nonumber\\
&\simeq& \frac{1}{2} g^2 \, \int d^d y \int d^dy' \,\varphi(y,0)
W^{(2)}(y,y') \varphi(y',0) \;,
\end{eqnarray}
where $W^{(2)}$ is the connected $2$-point function.
Finally, since we assume translation invariance along the mirrors' surface,
we conclude that:
\begin{equation}
\tilde{\lambda}(\kappa) \,=\, g^2 \, {\widetilde W^{(2)}}(\kappa) \;.
\end{equation}
Thus, for this first example, $\tilde{\lambda}$ may be read from the full
propagator for the material field on the mirror. In the relativistic case,
we know that:
\begin{equation}
    {\widetilde W^{(2)}}(\kappa)\,=\,\frac{z(\kappa)}{\kappa^2 + \Pi(\kappa)}
\end{equation}
where $z$ and $\Pi$ denote the wave function renormalization function and
the full self-energy, respectively. Of course, we assume that the $2$-point
function has been renormalized, so that there is no remaining ambiguity in
either one of these functions, which depend only on the $d$-scalar
$\kappa$. Since the full propagator may be related (via analytic
continuation) to response functions, we see that the renormalization conditions
(when needed) can be naturally identified with the {\em static\/} response
functions for the field, due to the microscopic degrees of freedom on the
mirrors.

Thus, unless quantum effects introduced qualitatively important changes in
the form of the full propagator, the conclusion is that, in this example, one
should expect to have a good description by using the free propagator,
namely,
\begin{equation}
\tilde{\lambda}(\kappa) \, \sim \,\frac{g^2}{\kappa^2 + m^2} \;,
\end{equation}
with the obvious changes if the matter field were described by a different
dispersion relation.

The zero-point energy density ${\mathcal E}_0$ can be written as
\begin{equation}\label{eq:defz}
{\mathcal E}_0 \;\equiv\; \lim_{T,L\to \infty} \frac{1}{L^{d-1} T}
\, \ln \big(\frac{\mathcal Z}{{\mathcal Z}_0}\big) \;,
\end{equation}
where
\begin{equation}
\frac{\mathcal Z}{{\mathcal Z}_0}=\frac{\int {\mathcal D\varphi}
e^{-S(\varphi)} }{\int {\mathcal D\varphi} e^{-S_0(\varphi)}}\;.
\end{equation}
This expression for the vacuum energy is of course equivalent to
the sum over eigenfrequencies of the full quantum system. For the
case of two mirrors, one is interested not in ${\mathcal E}_0$,
but rather in the subtracted quantity, $\tilde{\mathcal E}_0$,
defined as the difference
\begin{equation}
\tilde{\mathcal E}_0(a) \;\equiv\; {\mathcal E}_0 \,-\,{\mathcal
E}_0 (\infty)
\end{equation}
where ${\mathcal E}_0 (\infty)$ denotes the surface energy density
when the mirrors are separated by an infinite distance. Using the
results of Ref.\cite{flm08}, we can write the general expression
for the Casimir energy:
\begin{equation}
\tilde{\mathcal E}_0(a) \;=\; \frac{1}{2^d \pi^{d/2} \Gamma(d/2)} \,
\int_0^\infty d\kappa \, \kappa^{d-1} \; \ln \Big\{ 1 \,-\, \big[ 1
\, + \, \frac{2\kappa}{\tilde{\lambda}(\kappa)}\big]^{-2} \, e^{-2
\kappa a}\Big\} \;. \label{euclidresult}\end{equation}

The standard behaviour for $\tilde{\lambda}(\kappa)$ is $\sim \, \kappa^{-2}$
for large values of $\kappa$; this implies that the UV behaviour shall be
softened with respect to the perfect mirror case, as expected.  The IR
region of the integrand is, on the other hand, strongly affected by the
inclusion of the $2$-point function; this is more evident for the case of a
massless microscopic field.

Let us now consider the case of an Abelian gauge field $A_\mu$ in $d+1$
dimensions, whose free action is of the usual Maxwell type
\begin{equation}
S_0(A)\;=\; \int d^{d+1}x \frac{1}{4} F_{\mu\nu} F_{\mu\nu} \;.
\end{equation}
The coupling to the charged fields on the mirror, whose (total) electric
current is $J_a(y)$ will be assumed to be of the standard $J_a A_a$ type,
where the current component $J_d$ does not appear (we assume that no charge
can escape from the mirror).  Under these assumptions, the effective action
can be written in the following fashion:
\begin{equation}\label{eq:deseffg}
e^{-S_{eff}(A)} \;=\; \frac{\int {\mathcal D}\mu \, e^{-S_m(\mu)
\,+\, i e \int d^dy \, J_a(y) \, A_a(y,0)}}{\int {\mathcal D}\mu \,
e^{-S_m(\mu)}} \;,
\end{equation}
where we use ``$\mu$'' to denote an unspecified field (or fields) which
carry a conserved current, and $e$ is the coupling constant.  It is
straightforward to verify that this defines a gauge-invariant functional of
$A_a$: $S_{eff}(A)=S_{eff}(A+\partial\omega)$,   for every smooth function
$\omega$, as a consequence of current conservation.

Then, in the quadratic approximation for $S_{eff}(A)$,
\begin{equation}
    S_{eff}(A) \,\simeq \, \frac{1}{2} \,e^2\, \int d^dy \int d^dy' \,
    A_a(y,0) \, \Pi_{ab}(y,y') \, A_b(y',0) \;,
\end{equation}
the kernel $ \Pi_{ab}$ is given by the current-current correlation
function:
\begin{equation}
\Pi_{ab}(y,y') \;=\; \langle J_a(y,0)  J_b(y',0) \rangle \;,
\end{equation}
where the average symbols correspond to matter fields functional averaging,
\begin{equation}
\langle \ldots \rangle \;=\; \frac{\int {\mathcal D}\mu \,
\ldots\, e^{-S_m(\mu)}}{\int {\mathcal D}\mu \, e^{-S_m(\mu)}} \;.
\end{equation}
Gauge invariance implies that $\Pi_{ab}$ is transverse; this is more
easily formulated in Fourier space: $k_a {\widetilde \Pi}_{ab}(k) = 0$.

When the matter fields action is relativistic, we may write ${\widetilde \Pi}_{ab}$ in
terms of just a scalar function of $\kappa$,
\begin{equation}
{\widetilde \Pi}_{ab}(k) \;=\; {\widetilde \Pi}(\kappa) \,\kappa^2 \,
\delta^\perp_{ab}(k) \;,
\end{equation}
where $\delta^\perp_{ab}(k)$ is the transverse tensor:
\mbox{$\delta^\perp_{ab}(k)= \delta_{ab} - k_a k_b/\kappa^2$}.
With this information, we can then produce  more convenient expressions for
$S_{eff}(A)$,
\begin{eqnarray}
    S_{eff}(A) &\simeq& \frac{1}{2} \,e^2\, \int
    \frac{d^dk}{(2\pi)^d} {\tilde A}^*_a(k,0) \, {\widetilde
    \Pi}_{ab}(k) \,{\tilde A}_a(k,0) \nonumber\\
&=& \frac{1}{4} \,e^2\, \int
\frac{d^dk}{(2\pi)^d} {\tilde F}^*_{ab}(k,0) \, {\widetilde \Pi}(\kappa)
\,{\tilde F}_{ab}(k,0) \;,
\end{eqnarray}
which shows that $\kappa^2 {\tilde \Pi}$  plays a similar role to ${\tilde
\lambda}$ for the real scalar field.

When matter is {\em nonrelativistic\/}, the corresponding effective
action, $S^{nr}_{eff}$, will have the same structure as in the first line
of the previous equation,
\begin{equation}
    S^{nr}_{eff}(A) \;=\; \frac{1}{2} \,e^2\,
\int \frac{d^dk}{(2\pi)^d} {\tilde A}^*_a(k,0) \, {\widetilde
    \Pi}^{nr}_{ab}(k) \,{\tilde A}_a(k,0)
\end{equation}
with an, in principle, different vacuum polarization tensor ${\widetilde
\Pi}^{nr}_{ab}$. This tensor will still be transverse (current
conservation) but it will fail to be just a scalar times the transverse
$\delta$. Lack of relativistic invariance means that its structure will be
more complex, depending on {\em two\/} independent functions, which in turn
will depend on $|k_0|$ and $|{\mathbf k}|$ separately (note that
$|{\mathbf k}|$ is the norm of a $d-1$-dimensional vector).

Assuming rotation invariance on the $(d-1)$-dimensional spatial plane of
each mirror, we can decompose ${\widetilde \Pi}^{nr}_{ab}$, using now {\em
two\/} independent tensors:
$P^\perp_{ab}$ and $P^\parallel_{ab}$, that can be constructed having in
mind that only the Galilean group of symmetries is relevant here. A
convenient choice for those tensors is such that $P^\perp_{ab}$ is
transverse in the $d-1$-dimensional sense, namely,
\begin{equation}
    P^\perp_{00} \;=\;  0 \;=\; P^\perp_{0i}\;=\; P^\perp_{i0} \;,
\;\;\; P_{ij}^\perp \,=\, \delta_{ij} - \frac{k_i k_j}{|{\mathbf k}|} \;,
\end{equation}
where $i,\,j$ denote run over the values $1,\ldots, d-1$, and
\begin{equation}
    P^\parallel_{ab} \;=\;  \delta_{ab}^\perp \,-\,  P^\perp_{ab}
\;.
\end{equation}
Note that, with the conventions above, they verify:
\begin{equation}
 (P^\perp)^2 \,=\, P^\perp \;,\;\; (P^\parallel)^2 \,=\, P^\parallel
 \;,\;\; P^\perp P^\parallel = 0 \;,\;\;  P^\perp + P^\parallel =
\delta^\perp \;,
\end{equation}
and, on the other hand, each one of them is transverse to $k_a$.

Then the general structure of ${\widetilde
\Pi}^{nr}_{ab}$ is:
\begin{equation}
 {\widetilde \Pi}^{nr}_{ab} \,=\, {\widetilde \Pi}^\perp \, P^\perp_{ab}
\,+\, {\widetilde \Pi}^\parallel \, P^\parallel_{ab} \;,
\end{equation}
where ${\widetilde \Pi}^\perp$ and ${\widetilde \Pi}^\parallel$ are
functions of $|k_0|$ and $|{\mathbf k}|$.
These two functions are related to different properties of the medium. In
the $T>0$ case, for example, the function ${\widetilde \Pi}^\parallel$ is
responsible for the Debye screening of static electric fields, with a
screening length determined by ${\widetilde \Pi}^\parallel(k_0=0, |{\mathbf
k}| \to 0)$.  Static magnetic field, on the other hand, are not screened,
although there is non-vanishing screening when time-dependent magnetic
fields are considered.
They are also related to plasma oscillations, with plasma frequencies
determined also by the coefficients above.

Finally, to find the Casimir energy corresponding to the gauge field case,
we recall that that energy can be derived from the Euclidean functional:
\begin{equation}
{\mathcal Z} \;=\; \int \big[{\mathcal D}A\big] \, e^{-S(A)}
\end{equation}
where $\big[{\mathcal D}A\big]$ denotes the integration measure including
gauge fixing and, as in the scalar case, $S$ is the sum of the free action plus the
effective actions concentrated on the two mirrors:
\begin{eqnarray}
S_{eff}^{(I)} &=& \frac{1}{2} \,e^2\, \int dx_d \int dx_d' \,
\int \frac{d^dk}{(2\pi)^d} \; {\tilde A}^*_a(k,x_d) \,
{\widetilde \Pi}(\kappa) \,\kappa^2\, \nonumber\\
&\times &  \delta^\perp_{ab}(k)
\sum_{\alpha=1}^2 \delta(x_d-a_\alpha) \delta(a_\alpha - x'_d)
 \,{\tilde A}_b(k,x'_d) \;,
\end{eqnarray}
where we assumed relativistic matter fields.

The path integral must be gauge-fixed; we adopt the `axial' gauge
$A_d\equiv 0$. This is a convenient choice, since it yields a
quite simple form for the action:
\begin{equation}
S \;=\; \frac{1}{2} \,\int dx_d \int dx_d' \, \int
\frac{d^dk}{(2\pi)^d}\;  {\tilde A}^*_a(k,x_d) \,
M_{ab}(k;x_d,x'_d) \, {\tilde A}_b(k,x'_d) \;,
\end{equation}
where
$$M_{ab}(k;x_d,x'_d) \;=\; -\partial_d^2 \,
\delta(x_d-x'_d)\, \delta^\parallel_{ab}(k)$$
\begin{equation}
\,+\, \Big\{ \big[ ( -\partial_d^2 + \kappa^2) \, \delta(x_d-x'_d)
\,+\, e^2 \, \kappa^2 \, {\widetilde \Pi}(\kappa) \,
\big[\sum_{\alpha=1}^2 \delta(x_d-a_\alpha) \delta(a_\alpha - x'_d)\big] \Big\}
\delta^\perp_{ab}(k)
\end{equation}
and $\delta^\parallel_{ab}  \equiv \delta^\perp_{ab} - \delta_{ab}$.

The formal result for ${\mathcal Z}$ is
\begin{equation}
{\mathcal Z} \;=\; \Big( \det\big[ M_{ab}(k;x_d,x'_d) \big] \Big)^{-\frac{1}{2}}
\;,
\end{equation}
where the determinant is evaluated over all the (continuum and discrete)
indices.
Since each variable in the path integral may be uniquely decomposed into transverse
and longitudinal components: ${\tilde A}_a={\tilde A}^\perp_a + {\tilde
A}^\parallel_a$, we have that:
\begin{equation}
{\mathcal D}{\tilde A}\;=\; {\mathcal D}{\tilde A}^\perp
\; {\mathcal D}{\tilde A}^\parallel \;,
\end{equation}
and, as a consequence,
\begin{equation}\label{eq:decom}
{\mathcal Z} \;=\;
\Big( \det\big[ M^\perp(k;x_d,x'_d) \big] \Big)^{-\frac{d-1}{2}} \;
\Big( \det\big[ M^\parallel(k;x_d,x'_d) \big] \Big)^{-\frac{1}{2}}
\;,
\end{equation}
where
\begin{eqnarray}
M^\perp(k;x_d,x'_d) &=&  (-\partial_d^2 + \kappa^2) \, \delta(x_d-x'_d)
\nonumber\\
&+& e^2 \, \kappa^2 \, {\widetilde \Pi}(\kappa) \,
\big[\sum_{\alpha=1}^2 \delta(x_d-a_\alpha) \delta(a_\alpha - x'_d)\big] \nonumber\\
M^\parallel(k;x_d,x'_d) &=&  (-\partial_d^2) \,
\delta(x_d-x'_d)\;.
\end{eqnarray}
Note the power of $d-1$ in (\ref{eq:decom}), which appears as a
because $\delta^\perp_{ab}$ is the identity on the space of transverse
fields, a $d-1$ dimensional space.

From the previous expressions, we see that ${\mathcal Z}$ is equivalent to
the product of the vacuum functional for a massless free field which sees
no mirrors, times $d-1$ identical factors corresponding to scalar fields which
interact with the walls with a (common) coupling
${\tilde\lambda} = e^2 \,\kappa^2 \, {\tilde\Pi}(\kappa)$\;. The free scalar
field factor is irrelevant to the Casimir energy, since we subtract the
energy when the mirrors are infinitely distant, while the others yields
$(d-1)$ times a scalar field like energy
\begin{equation}
\tilde{\mathcal E}_0(a) \;=\;
\frac{d -1}{2^d \pi^{d/2} \Gamma(d/2)} \,
\int_0^\infty d\kappa \, \kappa^{d-1} \;
\ln \Big\{ 1 \,-\,
\big[ 1 \,+\, \frac{2}{e^2 \, \kappa {\Pi}(\kappa)}\big]^{-2}
\, e^{-2 \kappa a}\Big\} \;.
\end{equation}

We specialize the previous study to an interesting concrete example: that
of $d=3$ with matter described by a  massless Dirac field in a reducible $4
n$ component representation (which preserves parity), we have~\cite{bfo}:
\begin{equation}
{\widetilde \Pi}(\kappa) \;= \; \frac{n}{8 \kappa} \;,
\label{pi}
\end{equation}
so that, after introducing the dimensionless variable $x=\kappa a\,$,
\begin{equation}
\tilde{\mathcal E}_0(a) \;=\;
\frac{1}{8 \pi a^3} \,
\int_0^\infty dx \, x^2 \;
\ln \Big\{ 1 \,-\,
\big[ 1 \,+\, \frac{16}{e^2 n}]^{-2}
\, e^{-2 x}\Big\}=-\frac{C(e^2 n)}{a^3} \;.
\end{equation}
Incidentally, this is one of the cases studied in~\cite{flm08}.
This expression has the remarkable feature that the dependence of
the Casimir energy with the distance is exactly equal to the one
for a pair of perfect mirrors, albeit with a different global
factor $C(e^2 n)$ that interpolates  between zero, for vanishing
coupling constant $e^2=0$, and  the result for perfect mirrors for
$e^2\rightarrow\infty$.

This is to be contrasted to the non-relativistic case which, using the same gauge
fixing as before, yields a result composed of two contributions,
corresponding to the modes that are {\em spatially\/} transverse or
longitudinal:
$$
\tilde{\mathcal E}_0(a) \,=\,
\frac{d -2}{2^d \pi^{d/2} \Gamma(d/2)} \,
\int_0^\infty d\kappa \, \kappa^{d-1} \;
\ln \Big\{ 1 \,-\,
\big[ 1 \,+\, \frac{2}{e^2 \, \kappa {\widetilde
\Pi}^\perp(\kappa)}\big]^{-2}
\, e^{-2 \kappa a}\Big\}
$$
\begin{equation}
+ \,\frac{1}{2^d \pi^{d/2} \Gamma(d/2)} \,
\int_0^\infty d\kappa \, \kappa^{d-1} \;
\ln \Big\{ 1 \,-\,
\big[ 1 \,+\, \frac{2}{e^2 \, \kappa {\widetilde
\Pi}^\parallel (\kappa)}\big]^{-2}
\, e^{-2 \kappa a}\Big\} \;.
\end{equation}

%%%%%%%%%%%%%%%%%%%%%%%%%%%%%%%%%%%%%%%%%%%%%%%%%%%%%%%%%%%%%%%%%%%%%%%%%%
%%%%%%%%%%%%%%%%%%%%%%%%%%%%%%%%%%%%%%%%%%%%%%%%%%%%%%%%%%%%%%%%%%%%%%%%%%
%%%%%%%%%%%%%%%%%%%%%%%%%%%%%%%%%%%%%%%%%%%%%%%%%%%%%%%%%%%%%%%%%%%%%%%%%%
%%%%%%%%%%%%%%%%%%%%%%%%%%%%%%%%%%%%%%%%%%%%%%%%%%%%%%%%%%%%%%%%%%%%%%%%%%
%%%%%%%%%%%%%%%%%%%%%%%%%%%%%%%%%%%%%%%%%%%%%%%%%%%%%%%%%%%%%%%%%%%%%%%%%%

Finally, we attempt to reinterpret the previous results for the
Casimir energy as emerging from the existence of a discrete set of
modes, reflecting the existence of (imperfect) boundary conditions
at $x_d=0,a$. We shall focus, for the sake of simplicity, on the
case of a massless scalar vacuum field in $d+1$ dimensions. After
integrating out the matter degrees of freedom, one can evaluate
the in-in, Schwinger-Keldysh, or closed time path effective action
$S_{\rm eff}^{\rm CTP}$ \cite{calhu}. On general grounds, this
effective action has non-local kernels, denoting the existence of
dissipative and noise effects. Taking the functional variation of
the effective action, one can write a Langevin equation for the
system field as
\begin{equation}
\Box \varphi + g^2\sum_\alpha \int  dt' d^{d-1}x'_\parallel \;
G_{\rm ret}(t, {\mathbf x}_\parallel, t', {\mathbf x}'_\parallel)
\delta (x_d - x_\alpha) \varphi (t',
 {\mathbf x}'_\parallel ,x_\alpha) = \sum_\alpha K (t, {\mathbf x}_\parallel) \delta (x_d - x_\alpha) .
\label{eqmotion} \end{equation} where $G_{\rm ret}$ is the
dissipation kernel, and $K$ is a stochastic force whose
correlation function is related to $G_{\rm ret}$ via the
fluctuation-dissipation relation.

If we neglect the effects of the noise, the dissipative system will have complex  eigenfrequencies. In order to compute them,
we put the thin mirrors inside a larger
box of length $L$, and impose periodic boundary
conditions. Setting $K=0$ in Eq.(\ref{eqmotion}), it is possible to read the discontinuity of the
field $\varphi$ around the defect at $ x_d = 0$ and $x_d = a$ as
\begin{equation}
\varphi'\vert_{x_\alpha^+}-\varphi'\vert_{x_\alpha^-}
=g^2\sum_\alpha \int  dt' d^{d-1} x'_\parallel\;  G_{\rm ret}(t,
{\mathbf x}_\parallel, t', {\mathbf x}'_\parallel) \varphi (t',
{\mathbf x}'_\parallel , x_\alpha),
\end{equation}
with $\alpha = 0, a$.

The condition that defines the eigenfrequencies  is $f(\omega, a, L)=0$, with
\begin{equation}
f(\omega, a, L) = \frac{1}{4}\left(e^{4 i k_d L} - e^{2 i k_d
a}\right) \left(e^{-2ik_d a} - 1 \right) -
\frac{i k_d}{\lambda_{\rm ret}} \left(e^{4 ik_d L} - 1\right)+
\left(\frac{k_d}{\lambda_{\rm ret}}\right)^2\left(e^{2ik_d L}-
1\right)^2= 0, \label{modes}
\end{equation}
where $k_d=\sqrt{\omega^2- {\mathbf k}_\parallel^2}$ and $\lambda_{\rm ret} = \lambda_{\rm ret} (\omega, {\mathbf k}_\parallel)$ is the Fourier transform of the
retarded Green function $G_{\rm ret}$.  Note that  $\lambda_{\rm ret}$ is the appropriate analytic
continuation  of the Euclidean function
$\tilde\lambda(\kappa)$ to Minkowski space.

Let us first consider the case in which $\lambda_{\rm ret}$  is
real. When  $\lambda_{\rm ret}$ is constant, it corresponds to a
non dissipative situation, i.e. a free field with massive terms
concentrated on the position of the  mirrors \cite{milton}. On the
other hand, when $\lambda_{\rm ret}$ depends on frequency, it is
unrealistic to assume that it does not have an imaginary part,
because of causality. In spite of this, one can compute the
Casimir energy following a standard procedure. From Eq.
(\ref{modes}), it is easy to prove that, in the limit
$L\rightarrow\infty$, the admitted eigenfrequencies in the cavity
are real. In this particular case the Casimir energy can be
obtained as a sum-over-modes,
\begin{equation}
{\cal E} = \frac{1}{2}\sum_i \left(\omega_{0,i} - \omega_{\infty
,i}\right), \label{overmodes}\end{equation} which may be computed
using the so-called generalized argument theorem
\begin{equation}
{\cal E} = \frac{1}{4\pi i}\int\frac{d^{d-1}k_\parallel}{(2\pi)^{d-1}} \oint dz ~z
~\frac{d}{dz}~\ln \frac{f(z, a,
L)}{f(z,L/2,L)}.\label{argument}\end{equation}
The integral in the complex plane reduces to an integral on the imaginary
axis $z=iy$, and therefore, when $\lambda_{\rm ret}$ is an even
function of $\omega$, one gets
\begin{equation}
{\cal E}(a) = \frac{1}{2\pi} \int\frac{d^{d-1}k_\parallel}{(2\pi)^{d-1}} \int_0^\infty dy \ln\left\{1 -
e^{-2\sqrt{y^2+{\mathbf k}_\parallel^2} a}\left(1 + \frac{2\sqrt{y^2+{\mathbf k}_\parallel^2} }{\lambda_{\rm ret}
(i y, {\mathbf k}_\parallel)}\right)^{-2}\right\},
\label{lambdaconst}
\end{equation}
which is consistent with the result obtained in Eq.(\ref{euclidresult})
from the Euclidean effective action.

However, when the effective strength of the coupling between field
and mirrors comes from virtual effects of a matter field, one
expects to have a  {\em complex\/} spectrum.
For example, for the case of massless Dirac fields in $d=3$,
the Euclidean function $\tilde\lambda$ is given by
$\tilde\lambda(\kappa)=ne^2\kappa/8$ (see Eq.(\ref{pi})). Therefore,
the retarded analytic continuation to Minkowski spacetime is  \cite{bg}
\begin{equation}
\lambda_{\rm ret}(\omega,{\mathbf k}_\parallel) = -\frac{ne^2}{8}\left (
i\,\, {\rm sign} (\omega)\sqrt{\omega^2- {\mathbf k}_\parallel^2}\theta(\omega^2- {\mathbf k}_\parallel^2)
- \sqrt{{\mathbf k}_\parallel^2 - \omega^2}\theta({\mathbf k}_\parallel^2 -\omega^2)
\right )\,\, ,
\end {equation}
introducing a dissipative term into Eq. (\ref{eqmotion}).  It is
possible to show  that, in this particular case,  Eq.
(\ref{modes}) only admits propagating ($\omega^2> {\mathbf
k}_\parallel^2$) complex solutions. In the limit $L\rightarrow
\infty$, they are given by
\begin{equation}
\omega_m ^2= \left(\frac{m\pi}{a} - \frac{i}{a} \ln\vert 1 -
\frac{16}{ne^2}\vert\right)^2+{\mathbf k}_\parallel^2.\label{complexmodes}
\end{equation}
where $m$ is a natural number.

As expected, in order to evaluate the Casimir interaction energy
it is not correct to use the sum-over-modes formula  Eq.
(\ref{overmodes}), since the modes of the effective theory have
complex frequencies  \cite{henint}. The correct procedure is to
consider the sum over the real eigenfrequencies of the full
system, as implicitly done in the Euclidean approach.
Alternatively, one should include the effects of noise in the
effective theory, as done by Lifshitz in the derivation of the
well known formula for the Casimir interaction in the presence of
real media \cite{lifshitz}. However, it is remarkable that the
correct result derived using the Euclidean approach
Eq.(\ref{euclidresult}) formally coincides with the one derived
from the sum over eigenfrequencies assuming an unphysical
$\lambda_{\rm ret}$ which is real and non constant
Eq.(\ref{lambdaconst}). A similar situation holds for Lifshitz
formula, that has been derived using a sum over modes for
unphysical permittivities with no imaginary parts \cite{ded lif},
and also for the Van der Walls interaction between atoms immersed
in an absorptive medium \cite{sdm}.

%%%%%%%%%%%%%%%%%%%%%%%%%%%%%%%%%%%%%%%%%%%%%%%%%%%%%%%%%%%%%%%%%%%%%%%%%%
%%%%%%%%%%%%%%%%%%%%%%%%%%%%%%%%%%%%%%%%%%%%%%%%%%%%%%%%%%%%%%%%%%%%%%%%%%
%%%%%%%%%%%%%%%%%%%%%%%%% Conclusions %%%%%%%%%%%%%%%%%%%%%%%%%%%%%%%%%%%%
%%%%%%%%%%%%%%%%%%%%%%%%%%%%%%%%%%%%%%%%%%%%%%%%%%%%%%%%%%%%%%%%%%%%%%%%%%
%%%%%%%%%%%%%%%%%%%%%%%%%%%%%%%%%%%%%%%%%%%%%%%%%%%%%%%%%%%%%%%%%%%%%%%%%%
We conclude with some comments about the
previous models, the assumptions made to motivate them, and the  results
obtained therefrom.  An important point has to do with the relationship between the symmetries
of the model and the resulting (approximate) boundary conditions. We have
seen that, assuming relativistic invariance and parity conservation on the
mirrors' world volume, gauge invariance implies that the physics depends on
only one function. This function, the scalar part of the vacuum
polarization function, is only effective to impose a particular kind of
boundary condition, which affects the components of the electric field that
are parallel to the mirrors.  Had one wanted to impose more general
boundary conditions, one should have had to relax some assumptions
regarding symmetries.  This has been, indeed, explicitly shown for the case
of a nonrelativistic system, where there appear {\em two\/} different
functions. This situation also arises, for example, when the mirrors have a
finite width.
Finally, parity symmetry could have been broken, for example,
by considering mirrors which are permeated by a strong magnetic field
pointing along the direction of the $x_d$ coordinate. This set up would
produce exotic boundary conditions, mixing the transverse electric field
with the normal {\em magnetic\/} field.

\section*{Acknowledgements}
C.D.F. thanks CONICET, ANPCyT and UNCuyo for financial support. The work of
F.D.M. and F.C.L was supported by UBA, CONICET and ANPCyT.
We would like to thank D. Dalvit and P. Milonni for useful discussions.

%%%%%%%%%%%%%%%%%%%%%%%%%%%%%%%%%%%%%%%%%%%%%%%%%%%%%%%%%%%%%%%%%%%%%%
%%%%%%%%%%%%%%%%%%%%%%%% References %%%%%%%%%%%%%%%%%%%%%%%%%%%%%%%%%%
%%%%%%%%%%%%%%%%%%%%%%%%%%%%%%%%%%%%%%%%%%%%%%%%%%%%%%%%%%%%%%%%%%%%%%

\end{document}